\documentclass[journal,twoside]{IEEEtran}

\usepackage{cite}
\usepackage{mdwmath}
\usepackage{epsfig}
\usepackage{easymat}
\usepackage{amsfonts}
\usepackage{amsmath}
\usepackage{float}
\usepackage{cite}
\usepackage{graphicx}
\usepackage{balance}
\usepackage{subfigure}
\usepackage{latexsym}
\usepackage{amssymb}
\usepackage{txfonts}
\usepackage{wasysym}
\usepackage{mathbbol}
\usepackage{bm}
\usepackage{url}
\usepackage{flushend}

\newtheorem{theorem}{Theorem}
\newtheorem{corollary}{Corollary}

\DeclareMathOperator{\tr}{\mathrm{\textit{tr}}}

\newcommand{\avr}{\mathbb{E}}

\DeclareMathOperator{\Q}{\mathrm{\textit{Q}}}

\newcounter{mytempeqncnt}

\begin{document}

\title{Secure Diversity-Multiplexing Tradeoff of Zero-Forcing Transmit Scheme at Finite-SNR 
\author{ Zouheir Rezki~\IEEEmembership{Member,~IEEE,}
 and Mohamed-Slim Alouini,~\IEEEmembership{Fellow,~IEEE,}}
\thanks{Manuscript received November 13, 2010; revised August 20, 2011; accepted December 06, 2011.
 The editor coordinating the review of this paper and approving it for publication was Dr. David Love.}
\thanks{Zouheir Rezki and Mohamed-Slim Alouini are with the Electrical Engineering Program, Physical Science and Engineering (PSE) Division, King Abdullah University of Science and Technology (KAUST), Thuwal, Mekkah Province, Saudi Arabia, Email: \{zouheir.rezki,slim.alouini\}@kaust.edu.sa}
\thanks{Part of this work has been presented at the 2011 IEEE International Conference on Communications (ICC'2011), Kyoto, Japan.}
\thanks{Digital Object Identifier ./.}}

\markboth{IEEE TRANSACTION ON COMMUNICATIONS,~Vol.~X, No.~X, XXX~2012}%
{Rezki and Alouini: Secure Diversity-Multiplexing Tradeoff of Zero-Forcing Transmit Scheme at Finite-SNR}
%


\maketitle

\begin{abstract}
In this paper, we address the finite Signal-to-Noise Ratio (SNR) Diversity-Multiplexing Tradeoff (DMT) of the Multiple Input Multiple Output (MIMO) wiretap channel, where a Zero-Forcing (ZF) transmit scheme, that intends to send the secret information in the orthogonal space of the eavesdropper channel, is used. First, we introduce the secrecy multiplexing gain at finite-SNR that generalizes the definition at high-SNR. Then, we provide upper and lower bounds on the outage probability under secrecy constraint, from which secrecy diversity gain estimates of ZF are derived. Through asymptotic analysis, we show that the upper bound underestimates the secrecy diversity gain, whereas the lower bound is tight at high-SNR, and thus its related diversity gain estimate is equal to the actual asymptotic secrecy diversity gain of the MIMO wiretap channel. 
\end{abstract}

\section{Introduction}\label{sec-intro}
The wiretap channel in which a source communicates with a
receiver through a discrete, memoryless channel (DMC) and a
wire-tapper observes the output of this channel via another DMC,
has been introduced by Wyner \cite{wyner}. In this seminal work, it has been shown that if the physical channel to the eavesdropper is noisier than the channel to the legitimate receiver, then there exists an encoding-decoding scheme such that reliable communication with perfect-secrecy is possible (without the use of any encryption key). Motivated by this positive result, many other authors
have recently addressed the impact of fading on secure
communications. For instance, the effect of fading on secure communication for
single-antenna wiretap and broadcast channels has been
studied in \cite{Gopala,Liang,Khisti} where the secrecy-capacity
along with the optimal power allocation and/or rate-adaptation
strategies at the source have been derived under different Channel
State Information (CSI) assumptions. The secrecy-capacity of a multiple-antenna wiretap channel, with fixed
channel matrices (that are hence known to the transmitter) has
been addressed recently in \cite{Khisti-isit,Li,Oggier-isit,khisti-2008,Oggier-2007,Liu2,Ekrem,Tekin,Bagherikaram}.\

While the ergodic capacity is meaningful in fast-fading regime, where the codeword length spans many coherence periods, 
it is unfortunately unachievable in slow-fading regime, where the channel remains constant during the transmission of the codeword \cite{Tse-book}. In communication under secrecy constraint, the need for outage arises when 1) the transmitter does not have full CSI about the main channel or the
eavesdropper's channel or 2) the transmitter has full CSI but is subject to delay constraints. Therefore, resorting to an outage analysis in order to derive performance 
limits of communication systems is the only option. To this end, it
has been shown in, e.g., \cite{Parada,Barros-isit,Bloch} that in a
quasi-static fading channel and in contrast to the Gaussian
channel, secure communication is possible even if the average
Signal-to-Noise Ratio (SNR) of the main channel is less than that
of the wire-tapper (or one of the wire-tappers in a multiple
eavesdroppers case as discussed in \cite{Wang}). Moreover, if a
high level of outage is to be tolerated, then the outage secrecy
rate of the fading channel can even be higher than the secrecy
capacity of the Gaussian wiretap channel for similar average SNR
levels.\

A more global performance measure of a Multiple Input Multiple Output (MIMO) channel potential at high-SNR 
is the diversity-multiplexing tradeoff (DMT) of Zheng and Tse \cite{Zheng03}. In a seminal work, they showed that in a MIMO channel, both the diversity gain and the multiplexing gain can be achieved with
an optimal tradeoff. This tradeoff is a high-SNR characterization of the maximum
diversity gain available at each multiplexing gain. Recently, the high-SNR secrecy DMT of a
MIMO wiretap channel has been investigated in \cite{Yuksel-J}, where it has been shown that when all terminals perfectly know all the channels, a Zero-Forcing (ZF) transmit scheme that intends to send the secret information in the orthogonal space of the eavesdropper channel, is asymptotically (at infinite SNR) DMT-optimal.\

At low to moderate SNR (typically $3-20$ dB), where no secrecy constraint is considered, it has been shown that the asymptotic DMT
is an optimistic upper bound on the finite-SNR DMT \cite{Nara}. Indeed, the achievable diversity gains at realistic SNRs are
significantly lower than the asymptotic values. From a communication theoretical point of view, the finite-SNR framework also emphasizes an important open problem: How to design communication schemes that are not only optimal at high-SNR, but more importantly at finite-SNR too. The aim of this paper is to analyze the finite-SNR DMT of ZF transmit scheme under secrecy constraint, where all terminals have perfect CSI. Section \ref{S2} presents the system model and introduces the related
definitions. In section \ref{S3}, outage analysis is provided. Upper and lower bounds on the secrecy outage probability of ZF transmit scheme and their respective diversity gain estimates are presented in section \ref{S4}. We then analyze asymptotic behavior of the derived results in section \ref{S5}. Results related to a Gaussian approximation of ZF performances are reported in section \ref{S6}. Numerical results
are reported in Section \ref{S7}. Section \ref{S8} concludes the paper.

\section{Channel Model and Related definitions }\label{S2}
We consider a Gaussian wiretap MIMO channel in which the source attends to communicate 
confidential messages $w$ using $N_{t}$ transmit antennas to the
destination with $N_{m}$ receive antennas in the presence of an
eavesdropper with $N_{e}$ receive antennas. The outputs
at both the legitimate destination and the eavesdropper, at symbol time index $i=1,\ldots,L$, are expressed, respectively by:
\begin{equation}\label{E21}
\begin{cases}
  \textbf{y}_{m}(i) = \textbf{H}_{m}{\textbf{x}}(i)+{\textbf{n}}_{m}(i) \\
  {\textbf{y}}_{e}(i) = {\textbf{H}}_{e}{\textbf{x}}(i)+{\textbf{n}}_{e}(i)
\end{cases}
\end{equation}
where ${\textbf{x}}(i) \in \mathbb{C}^{N_{t}}$ is the transmitted
signal, and ${\textbf{H}}_{m}\in \mathbb{C}^{N_{m} \times
N_{t}}$, ${\textbf{H}}_{e}\in \mathbb{C}^{N_{e} \times N_{t}}$
represent the main channel and the wiretap channel gains,
respectively; and ${\textbf{n}}_{m}(i)\in \mathbb{C}^{N_{m}}$,
${\textbf{n}}_{e}(i)\in \mathbb{C}^{N_{e}}$ are circularly symmetric
white Gaussian noises with covariance matrices $
\avr[{\textbf{n}}_{m}(i){\textbf{n}}_{m}(i)^{\dag}]={I}_{N_{m}}$
and $
\avr[{\textbf{n}}_{e}(i){\textbf{n}}_{e}(i)^{\dag}]={I}_{N_{e}}$.
We first assume that ${\textbf{H}}_{m}$ and
${\textbf{H}}_{e}$ are fixed and deterministic matrices and that they are both known at the transmitter and the eavesdropper. We also assume that ${\textbf{H}}_{m}$ is known at the legitimate receiver. The source is constrained according to an average power
constraint:
\begin{equation}\label{E22}
     \frac{1}{L}\sum_{i=1}^{L}{\avr{\left[\|\textbf{x}(i)\|^{2}\right]}} \leq \eta,
\end{equation}
where $\eta$ is the
mean SNR per channel use at each receive antenna. We assume that CSI is available at both receivers and at the transmitter as well. The level of
uncertainty about the message $w$ at the eavesdropper is
measured by the equivocation rate defined by:
\begin{equation}\label{E23}
    R_e \coloneqq \frac{1}{L}H({\textbf{w}}\mid
    {\textbf{y}}_{e}^{(L)})
\end{equation}
where $H(\textbf{w}\mid \textbf{y})$ is the conditional entropy of $\textbf{w}$ given $\textbf{y}$ and ${\textbf{y}}_{e}^{(L)}=({\textbf{y}}_{e}(1),\ldots,{\textbf{y}}_{e}(L))$. 
The higher the equivocation rate is, the smaller is the
knowledge of the eavesdropper about the source. We focus on a constant 
secrecy rate transmission for which $R=R_{e}$. A rate $R$ is an 
achievable secrecy rate if for all $\epsilon
> 0$, there exists an encoder-decoder $(2^{LR},L,P_{e})$ for which $
2^{LR}$ represents the number of messages to be sent to the destination, $R_{e} \geq R-\epsilon$ and $P_{e} \leq
\epsilon$, where $P_{e}$ is the average error probability defined by:
\begin{equation}\label{E24}
P_{e}=\frac{1}{2^{LR}}\sum_{w=1}^{2^{LR}}{\text{Pr}\{\textbf{w}\neq\hat{\textbf{w}}\}},
\end{equation}
where $\hat{\textbf{w}}$ is the output of the decoder at the intended receiver as a result of observing $\textbf{y}_{m}^{\left(L\right)}$. Furthermore, the secrecy capacity is given by:
$C_{s}\coloneqq
    \underset{R \in \mathcal{R}_{s}} \sup \ R$, where $\mathcal{R}_{s}$ 
is the set of achievable secrecy rates. The secrecy capacity of the MIMO wiretap channel in nats per channel use (npcu) is given by \cite{khisti-2008,Oggier-2007,Liu2}:
\begin{equation}\label{E25}
    C_{s} = \underset{\tr{\left(Q_{x}\right)}\leq \eta} \max{
    \biggl[\Psi{\left(Q_{x},H_{m}\right)} - \Psi{\left(Q_{x},H_{e}\right)}\biggr]^{+}},
\end{equation}
where $\Psi{(Q,H)}= \log{\det{\left(H Q H^{\dag}+ I\right)}}$. An achievability scheme of (5) is described as follows: Let $R_{s}=C_{s}-\delta$ be the secrecy rate, for an arbitrary small $\delta >0$. To transmit one of the $2^{LR_{s}}$ equally likely messages $w \in \left\{1,2,\ldots,2^{LR_{s}}\right\}$, the sender employs $2^{LR_{s}}$ codes, $\mathcal{C}_{w}$, $w=1,2,\ldots,2^{LR_{s}}$. Each code $\mathcal{C}_{w}$ contains $2^{L \left(\Psi{\left(Q_{x},H_{e}\right)}-\epsilon\right)}$ Gaussian codewords $x^{(L)}$ of length $L$ symbols drawn from $\mathcal{CN}\left(0,Q\right)$. To encode a message $w$, a codeword $x^{(L)}$ is chosen uniformly at random from $\mathcal{C}_{w}$ and is transmitted through the channel, using $N_{t}$ antennas. By a random coding argument, it can be shown that there exists at least one such a codebook with an arbitrary small $P_{e}$ and with the equivocation rate $R_{e}$ arbitrary close to $R_{s}$, as $L \rightarrow \infty$ \cite{wyner}, \cite[Section II.B]{Tang}.\

Next, we consider a quasi-static fading model in which ${\textbf{H}}_{m}$ and
${\textbf{H}}_{e}$ in (\ref{E21}) have now independent
and identically-distributed complex Gaussian entries with zero-mean and
unit-variance. The notions of secrecy capacity is still meaningful if the matrices are known
to all terminals. However, It loses significance when either the matrices are unknown or the transmitter
has to fix a target rate due to delay constraints. In the later case, it is clear that no matter how small the secrecy rate at which we want to communicate, there is a non zero probability that the pair $\left(\textbf{H}_{m},\textbf{H}_{e}\right)$ is not able to support it. The performance limit in this case may be captured through an outage analysis as argued in Section \ref{S3}.

For an SNR-dependent secrecy rate $R_{s}(SNR)$, the secrecy multiplexing and diversity gains are defined by: 
\begin{IEEEeqnarray}{rCl}
     r_{s} & = &  \frac{R_{s}}{\log(1+g \cdot \eta)} \label{E26} \\
     d_{s}(r_{s},\eta) & = & -\eta {\frac{\partial{\left(\log P_{e}(r_{s},\eta)\right)}}{\partial{
     \eta}}}, \label{E27}
\end{IEEEeqnarray}
where $g$ is an array gain that is chosen to guarantee a fair comparison of the outage performance with secrecy constraint, across different antenna configurations, at low to medium SNR values. Indeed, by considering the MIMO secrecy rate at low SNR, it has been shown in \cite{Gursoy} that at low SNR, the secrecy capacity can be expressed by:
\begin{equation}\label{E28}
C_{s}= \left[\lambda_{max}\left(\textbf{H}_{m}^{\dag}\textbf{H}_{m}-\textbf{H}_{e}^{\dag}\textbf{H}_{e}\right)\right]^{+} \cdot \eta + o(\eta),
\end{equation}
where $\lambda_{max}\left(\textbf{H}_{m}^{\dag}\textbf{H}_{m}-\textbf{H}_{e}^{\dag}\textbf{H}_{e}\right)$ is the maximum eigenvalue of $\textbf{H}_{m}^{\dag}\textbf{H}_{m}-\textbf{H}_{e}^{\dag}\textbf{H}_{e}$. Thus $g$ is chosen such that $g=\avr{\left\{\left[\lambda_{max}\left(\textbf{H}_{m}^{\dag}\textbf{H}_{m}-\textbf{H}_{e}^{\dag}\textbf{H}_{e}\right)\right]^{+}\right\}}$. Note that if $\lambda_{max}\left(\textbf{H}_{m}^{\dag}\textbf{H}_{m}-\textbf{H}_{e}^{\dag}\textbf{H}_{e}\right) \leq 0$, then $\textbf{H}_{m}^{\dag}\textbf{H}_{m}\preceq \textbf{H}_{e}^{\dag}\textbf{H}_{e}$ and thus the main channel is a degraded version of the eavesdropper one, which implies that the secrecy capacity is equal to zero and so is the secrecy multiplexing gain $r_{s}$ in (\ref{E26}). As defined in (\ref{E26}), the secrecy multiplexing gain $r_{s}$ provides an indication 
as to how the secrecy rate scales with the capacity 
of an Additive White Gaussian Noise (AWGN) channel with an array gain $g$, as the SNR changes. Note that for a constant secrecy multiplexing gain, the secrecy rate increases as the SNR increases. Hence, the secrecy multiplexing gain $r_{s}$
provides an indication of the sensitivity of a secure rate adaptation
strategy as the SNR changes. On the other hand, the secrecy diversity
gain $d_{s}(r_{s},\eta)$ of a secure rate adaptive
system, with a fixed multiplexing gain $r_{s}$ at SNR $\eta$, represents the negative slope of the log-log plot of the average error probability
versus SNR. It can then be used to estimate the additional SNR required to
decrease the average error probability $P_{e}\left(r_{s},\eta\right)$ by a specific amount for a given
multiplexing gain $r_{s}$. This interpretation of the diversity gain is meaningful no matter how $P_{e}$ scales with $\eta$. However, as the SNR $\eta$ tends toward infinity, the diversity gain in (\ref{E27}) coincides with the asymptotic (at high-SNR) one as defined in \cite{Zheng03}. Recall that the finite-SNR multiplexing and diversity gains have been first introduced in \cite{Nara_IT} to estimate the performance limits of multiple antenna communication systems, but without secrecy constraint. By properly extending the notion of finite-SNR DMT to secure communication, we intend to provide an estimate of its performance limits, not asymptotically \cite{Yuksel-J}, but at realistic SNR values.\

Finally, we consider perfect-secrecy communication in which we let the size of the sub-codebooks $C_{w}$ varies with the eavesdropper channel, i.e., $\left|C_{w}\right|=2^{L \  \Psi{\left(Q_{x},H_{e}\right)}}$. That is, to ensure a secrecy rate $R_{s}$, the transmission rate at the source is equal to:  
\begin{equation}\label{E29}
R=R_{s}+ \Psi{\left(Q_{x},H_{e}\right)}.
\end{equation}

\section{Outage Analysis and Related Background}\label{S3}
\subsection{Background}\label{S31}
Recall that in a quasi-static channel model without secrecy constraint, a detection error happens as a result of three events: 1) the channel matrix
is atypically ill-conditioned, 2) the additive noise is atypically
large, or 3) some codewords are atypically close together. For large coherence block $L$ (which we assume in this paper) and using capacity-achieving codes, the randomness in the last two events is averaged out, and thus we can focus only on the bad channel event. That is, the error probability without secrecy constraint is mainly dominated by the outage events \cite{Zheng03}. Now, introducing the secrecy constraint engenders a new error event at the legitimate receiver corresponding to the case where both the legitimate and the eavesdropper can decode the message correctly, and thus the secrecy is not achieved \cite{Thangaraj}. The later error event may be defined by:
\begin{equation*}
\text{P}\left\{\text{secrecy is not achieved}\right\}=\text{P}\left\{ \frac{1}{L}H({\textbf{w}}\mid
    {\textbf{y}}_{e}^{(L)}) < R_{s}\right\}.
\end{equation*}
Hence, in order to compute the error probability of the scheme, we only need to analyze the probability that the main channel is in outage or the secrecy is not achieved. That is:
\begin{equation}\label{ER1}
P_{e}=\text{P}\left\{\text{ main channel is in outage or secrecy is not achieved}\right\}.
\end{equation}
Note that (\ref{ER1}) is rather general and does not depend on the coding scheme used at the transmitter. Therefore, in order to analyze the finite-SNR
diversity-multiplexing tradeoff of the MIMO wiretap channel, a characterization of the main channel outage and ``secrecy is not achieved'' events is
required.\

When perfect CSI is available at all terminals,  the outage probability has been characterized at high-SNR in \cite{Yuksel-J}. The asymptotic secrecy DMT has been derived consequently. In particular, it has been shown that at high-SNR and under perfect CSI at all terminals, if $N_{e} < N_{t}$, then the secrecy DMT is a piecewise linear function joining the points $\left(l,d_{s}^{\infty}(l)\right)$, where $l=0,\ldots,\min{\left(N_{t}- N_{e},N_{m}\right)}$, with $d_{s}^{\infty}(l)$ given by:
\begin{equation}\label{E31}
d_{s}^{\infty}(l)=(N_{t}- N_{e}-l)(N_{m}-l).
\end{equation} 
If on the contrary $N_{e} \geq N_{t}$, then the secrecy DMT reduces to the single point $\left(0,0\right)$. From a DMT perspective, the MIMO wiretap channel is equivalent to a reduced MIMO channel with $\left(N_{t} - N_{e}\right)$ and $N_{m}$ transmit and receive antennas, respectively. As the error probability is generally log-concave \cite{Conti,Loyka}, the diversity gain is an increasing function in SNR. Therefore, the finite-SNR secrecy diversity gain $d_{s}$ is equal to zero if $N_{e} \geq N_{t}$. Next, we assume that $N_{e} < N_{t}$ and we let $m=\min{(N_{t}-N_{e},N_{m})}$ and $k=\max{(N_{t}-N_{e},N_{m})}$, for notation convenience.

\subsection{Outage Analysis}\label{S32}
Since in our scheme, to transmit messages securely, we have used a family of capacity-achieving codes over the wiretap channel (please see our achievability scheme above), then by \cite[Theorem 1]{Thangaraj}, the secrecy is guaranteed and (\ref{ER1}) becomes: 
\begin{equation}\label{ER2}
P_{e}=\text{P}\left\{\text{main channel is in outage}\right\}.
\end{equation}
An outage occurs if the actual channel realization $\left(\textbf{H}_{m},\textbf{H}_{e}\right)$ belongs to $\mathcal{O}_{1}$, defined by: 
$\mathcal{O}_{1}=\biggl \{\textbf{H}_{m}, \textbf{H}_{e} \mid \Psi{\left(Q_{x},\textbf{H}_{m}\right)} < R \biggr \}$. Using (\ref{E29}), the main channel outage probability may be expressed by:
\begin{equation}\label{E32}
P_{out}=\underset{Q_{x}\succeq 0, \tr{\left(Q_{x}\right)}\leq \eta}\inf{P\biggl[\Psi{\left(Q_{x},\textbf{H}_{m}\right)}-\Psi{\left(Q_{x},\textbf{H}_{e}\right)} < R_{s}\biggr]}.
\end{equation}
Since the structure of the optimal covariance matrix is generally not known for an arbitrary SNR, characterizing $P_{out}$ is still an insolvable problem. Nevertheless, any choice of $Q_{x}\succeq 0$ that satisfies the trace constraint, provides an upper bound on the outage probability. Motivated by the results at high-SNR in \cite{Yuksel-J} where it has been shown that a ZF transmit scheme is DMT-optimal, we investigate next the finite-SNR performance of the later scheme. Recall that a ZF transmit scheme consists of transmitting the secret information in the null space of the eavesdropper channel. That is, the transmitter sends a vector $\textbf{x}=A\textbf{u}$, where $A$ is such that $H_{e}^{\bot}=AA^{\dag}$, where $H_{e}^{\bot}$ is the orthogonal matrix of $\textbf{H}_{e}$, and where $\textbf{u}$ is a Gaussian vector with covariance $\frac{\eta}{N_{t}-N_{e}}I_{N_{t}-N_{e}}$. This choice of $\textbf{x}$ provides the following upper bound on the outage probability:
\begin{equation}\label{E33}
P_{out} \leq P_{out}^{ZF},
\end{equation}
where $ P_{out}^{ZF}=P\biggl[\Psi{\left(\textbf{H}_{eq}\right)}< R_{s}\biggr]$, with $\psi{\left(\textbf{H}_{eq}\right)}=\log{\det{\left(I+\frac{\eta}{N_{t}-N_{e}}\textbf{H}_{eq} \textbf{H}_{eq}^{\dag}\right)}}$ and with $\textbf{H}_{eq}=\left(\textbf{H}_{m}A\right)$ an $N_{m} \times \left(N_{t}-N_{e}\right)$ channel matrix with i.i.d. complex Gaussian entries with zero-mean and unit-variance. $P_{out}^{ZF}$ represents the outage probability of an equivalent MIMO channel with $\left(N_{t}-N_{e}\right)$ and $N_{m}$ transmit and receive antennas, respectively, and where the input is Gaussian with covariance $\frac{\eta}{N_{t}-N_{e}}I_{N_{t}-N_{e}}$. From \cite{Zheng03}, we know that the $P_{out}^{ZF}$ achieves the high-SNR DMT of this MIMO channel, which is equal to $d_{s}^{\infty}$ given by (\ref{E31}), and thus ZF is also asymptotically optimal with respect to the secrecy DMT. At finite-SNR, we expect ZF to induce a power loss that may translate to a decrease of its secrecy diversity gain.  

\section{ZF Outage Analysis At Finite-SNR}\label{S4}
In this section, upper and lower bounds on the secrecy outage probability of ZF transmit scheme are derived. In order to characterize the secrecy diversity gain of this transmission scheme, these bounds are then used to provide insightful diversity estimates.
\subsection{Upper Bound on ZF Secrecy Outage Probability }
Our results is summarized in the following theorem.
\begin{theorem}[Upper bound on $P_{out}^{ZF}$]\label{T1} 
An upper bound on the outage probability of ZF transmit scheme is given by:
\begin{equation}\label{E41}
P_{out}^{ZF} \leq \prod_{l=1}^{m}\Gamma_{inc}(\xi(r_{s}),k-l+1)\left(1-\prod_{l=1}^{m}\Bigl[1-\frac{\Gamma_{inc}(\xi(b_{l}),k-l+1)}{\Gamma_{inc}(\xi(r_{s}),k-l+1)}\Bigr]\right),
\end{equation}
where $\Gamma_{inc}$ is the
incomplete Gamma function defined by
$\Gamma_{inc}(x,a)=\frac{1}{(a-1)!}\int_{0}^{x}t^{a-1}e^{-t}dt$, where  $\xi\left(x\right)=\frac{N_t-N_{e}}{\eta}\left(\left(1+g\eta\right)^{x}-1\right)$ and where $b_{l}$, $l=1,\dots,m$, are arbitrary positive coefficients
that satisfy $0\leq b_{m} \leq \ldots \leq b_{1}$ along with $r_{s}=\sum_{l=1}^{m}b_l$.
\end{theorem}
\begin{IEEEproof}
Let $\textbf{H}_{eq}=Q\textbf{R}$ be the orthogonal triangular (QR) decomposition of the matrix $\textbf{H}_{m}A$, where $
Q$ is an $N_{m} \times N_{m}$ unitary matrix and where
$\textbf{R}$ is an $N_{m} \times \left(N_{t}-N_{e}\right)$ upper triangular matrix
with independent entries. The square magnitudes of the diagonal
entries of $\textbf{R}$, $|\textbf{R}_{l,l}|^2$, are chi-square
distributed with $2(k-l+1)$ degrees of freedom, $l=1,\ldots
,m$. The off-diagonal elements of $\textbf{R}$ are i.i.d. Gaussian variables, with zero mean and unit
variance. Using the fact that $\det(I+\textbf{R} \textbf{R}^{\dag}) \geq \prod_{l=1}^{m}\left(1+\left|\textbf{R}_{l,l}\right|^{2}\right)$, the following lower bound is obtained:
\begin{equation}\label{E42}
\Psi{\left(\textbf{H}_{eq}\right)} \geq \sum_{l=1}^{m}\log{\left(1+\frac{\eta}{N_{t}-N_{e}}\left|\textbf{R}_{l,l}\right|^{2}\right)}.
\end{equation}
Then $P_{out}^{ZF}$ can be upper bounded as follows:
\begin{IEEEeqnarray}{rCl}
P_{out}^{ZF} &\leq& P\biggl[ \sum_{l=1}^{m}\log{\left(1+\frac{\eta}{N_{t}-N_{e}}\left|\textbf{R}_{l,l}\right|^{2}\right)} < R_{s}\biggr]  \nonumber \\
              &=& P\biggl[ \log{\left(1+\frac{\eta}{N_{t}-N_{e}}\left|\textbf{R}_{l,l}\right|^{2}\right)} < R_{s}, l=1,\ldots,m\biggr] \nonumber \\
              & \quad& -P\biggl[ \sum_{l=1}^{m}\log{\left(1+\frac{\eta}{N_{t}-N_{e}}\left|\textbf{R}_{l,l}\right|^{2}\right)} \geq R_{s}; \nonumber \\       
              & \quad& \log{\left(1+\frac{\eta}{N_{t}-N_{e}}\left|\textbf{R}_{l,l}\right|^{2}\right)} \leq R_{s}, l=1,\ldots,m\biggr] \nonumber \\
              &\leq& P\biggl[ \log{\left(1+\frac{\eta}{N_{t}-N_{e}}\left|\textbf{R}_{l,l}\right|^{2}\right)} < R_{s}, l=1,\ldots,m\biggr] \nonumber \\
              &\quad& - P\biggl[ \log{\left(1+\frac{\eta}{N_{t}-N_{e}}\left|\textbf{R}_{l,l}\right|^{2}\right)} \leq R_{s}; \nonumber \\ 
              &\quad& b_{l} \log{\left(1+g \eta \right)} \leq \log{\left(1+\frac{\eta}{N_{t}-N_{e}}\left|\textbf{R}_{l,l}\right|^{2}\right)}, l=1,\ldots,m\biggr] \nonumber \\ 
              &=& \prod_{l=1}^{m}\Gamma_{inc}(\xi\left(r_{s}\right),k-l+1)  \nonumber \\
              &\quad& - \prod_{l=1}^{m}\biggl(\Gamma_{inc}(\xi\left(r_{s}\right),k-l+1)-\Gamma_{inc}(\xi\left(b_{l}\right),k-l+1)\biggr) \label{E43}
\end{IEEEeqnarray}
where (\ref{E43}) follows because $\left|R_{l,l}\right|^{2}$ is chi-square distributed with $2\left(k-l+1\right)$ degrees of freedom. From (\ref{E43}), (\ref{E41}) follows immediately.
\end{IEEEproof} 
The upper bound in (\ref{E41}) is minimized over all $b_{l}$, $l=1,\ldots,m$, that satisfy $0\leq b_{m} \leq \ldots \leq b_{1}$ along with $r_{s}=\sum_{l=1}^{m}b_l$ to obtain a better result. It turns out that this minimization problem is convex and can be easily solved using standard optimization algorithms. Furthermore, even without such an optimization, the upper bound in (\ref{E41}) involves $\Gamma_{inc}\left(x,N\right)$ functions that are now well studied and are built-in functions in most popular computing softwares. Moreover, since the asymptotic behavior of $\Gamma_{inc}\left(x,N\right)$ as $x \longrightarrow 0$, or $x \longrightarrow \infty$ is well understood, the upper bound in (\ref{E41}) is analytically tractable asymptotically. As evidence of our claim, we provide below some results related to this upper bound.
\begin{corollary}[Secrecy Diversity Gain Estimate $\hat{d}_{s}^{U}$]\label{C1}
Let $\alpha_{l}=\Gamma_{inc}(\xi(r_{s}),k-l+1)$, $\beta_{l}=\Gamma_{inc}(\xi(b_{l}),k-l+1)$ and $f_{l}(x)$ defined by:
\begin{equation}\label{E44}
f_{l}(x)=\left((1+g\eta)^{x}-x g \eta (1+g\eta)^{x
 -1}-1\right) \frac{\xi(x)^{k-l}e^{-\xi(x)}/(k-l)!}{{\Gamma_{inc}(\xi(x),k-l+1)}}.
\end{equation}
Then, a secrecy diversity gain estimate $\hat{d}_{s}^{U}$ (U for related to the upper bound on $P_{out}^{ZF}$) of ZF transmission scheme is given by:
\begin{equation}\label{E45}
\hat{d}_{s}^{U}(r_{s},\eta) = \frac{N_{t}-N_{e}}{\eta}\sum_{l=1}^{m}\Biggl[ f_{l}(r_{s})+\frac{\beta_{l}}{\alpha_{l}}\biggl(f_{l}(b_{l})-f_{l}(r_{s})\biggr)\frac{\underset{k\neq l}\prod\left(1-\frac{\beta_{k}}{\alpha_{k}}\right)}{1-\overset{m} {\underset{k=1} \prod}\left(1-\frac{\beta_{k}}{\alpha_{k}}\right)}\Biggr].
\end{equation}
\end{corollary}
\begin{IEEEproof}
The proof follows directly from (\ref{E41}) and (\ref{E27}).
\end{IEEEproof} 
The secrecy diversity gain estimate $\hat{d}_{s}^{U}(r_{s},\eta)$ in (\ref{E45}) is a finite sum that can be easily evaluated. 

\subsection{Lower Bound on ZF Secrecy Outage Probability }
Our result is stated in the following theorem.
\begin{theorem}[Lower Bound on $P_{out}^{ZF}$]\label{T2} 
A lower bound on the outage probability of ZF transmit scheme is given by:
\begin{equation}\label{E46}
P_{out}^{ZF} \geq \prod_{l=1}^{m}\Gamma_{inc}(\xi(a_{l}),k+m-2l+1),
\end{equation}
where $a_{l}$, $l=1,\dots,m$, are arbitrary positive coefficients
that satisfy $0\leq a_{m} \leq \ldots \leq a_{1}$ along with $r_{s}=\sum_{l=1}^{m}a_l$.
\end{theorem}
\begin{IEEEproof}
Similarly to the proof of Theorem \ref{T1}, and using the fact that $\det(\textbf{T})\leq \underset{l}\prod{T_{l,l}}$, for any
nonnegative-definite matrix $\textbf{T}$, the following upper bound is obtained:
\begin{equation}\label{E47}
\Psi{\left(\textbf{H}_{eq}\right)} \leq \sum_{l=1}^{m}\log{\left(1+\frac{\eta}{N_{t}-N_{e}} \bm{\Delta}_l \right)},
\end{equation}
where $\bm{\Delta}_l=\overset{m} {\underset{k=l}\sum}|{\textbf{R}_{l,k}}|^2$, $l=1,\ldots,m$,
is the $l^{th}$ diagonal entry of $\textbf{R} \textbf{R}^{\dag}$. 
Then $P_{out}^{ZF}$ can be lower bounded as follows:
\begin{IEEEeqnarray}{rCl}
P_{out}^{ZF} &\geq& P\biggl[ \prod_{l=1}^{m}\left(1+\frac{\eta}{N_{t}-N_{e}} \bm{\Delta}_{l}\right) < \left(1+g\eta\right)^{r_{s}}\biggr] \nonumber \\
             &\geq& P\biggl[(1+\frac{\eta}{N_{t}-N_{e}}\bm{\Delta}_l)<(1+g\eta)^{a_l},\quad l=1,\dots,m\biggr] \nonumber \\
             &=   & \prod_{l=1}^{m}Prob\left(\bm{\Delta}_l< \xi(a_l) \right) \nonumber,
\end{IEEEeqnarray}
from which the lower bound in (\ref{E46}) follows because $\bm{\Delta}_l$ is chi-square distributed with $2(k+m-2l+1)$ degrees of freedom.
\end{IEEEproof} 
The lower bound in (\ref{E46}) is then maximized over all $a_{l}$, $l=1,\ldots,m$, that satisfy $0\leq a_{m} \leq \ldots \leq a_{1}$ along with $r_{s}=\sum_{l=1}^{m}a_l$ to obtain a better result. 
\begin{corollary}[Secrecy Diversity Gain Estimate $\hat{d}_{s}^{L}$]\label{C2}
A secrecy diversity gain estimate $\hat{d}_{s}^{L}$ of ZF transmission scheme is given by:
\begin{equation}\label{E48}
\hat{d}_{s}^{L}(r_{s},\eta) = \frac{N_{t}-N_{e}}{\eta}\sum_{l=1}^{m} f_{2l-m}(a_{l})
\end{equation}
\end{corollary}
\begin{IEEEproof}
The proof follows directly from (\ref{E46}) and (\ref{E27}).
\end{IEEEproof} 
Note that the secrecy diversity gain estimate $\hat{d}_{s}^{L}(r_{s},\eta)$ in (\ref{E48}) is again a finite sum that can be easily evaluated. To conclude this section, few remarks are worthwhile:
\begin{enumerate}
	\item When $N_{t}-N_{e}=1$ or $N_{m}=1$, the upper bound (\ref{E41}) and the lower bound (\ref{E46}) coincide, the secrecy outage probability of ZF transmit scheme is completely characterized at all SNR values, and the secrecy diversity gain is equal to:
	\begin{equation}\label{E49}
	{d}_{s}(r_{s},\eta) = \frac{N_{t}-N_{e}}{\eta}f_{1}(r_{s}).
	\end{equation}
	\item From the proof of Theorem \ref{T2}, it is easy to see that $P_{out}^{ZF} \leq \prod_{l=1}^{m}\Gamma_{inc}(\xi(r_{s}),k-l+1)$. This ``naive'' upper bound is however loose as it fails to provide any positive diversity gain estimate for $r_{s} \geq 1$ (cf. Appendix \ref{App2} for details). The upper bound in (\ref{E41}) provides better results.
	\item By (\ref{E33}), the upper bound in (\ref{E41}) is also an upper bound on the outage probability of the MIMO wiretap channel.
\end{enumerate}

\section{Asymptotic Analysis}\label{S5}             
In this section, we analyze the derived results in section \ref{S4}, at low secrecy multiplexing gain regime ($r_s \rightarrow 0$), which captures the maximum secrecy diversity gain estimates for a given SNR value, and at high-SNR regime ($\eta \rightarrow \infty.$), which characterizes the asymptotic DMT achieved by the upper and the lower bounds reported in Theorem 1 and Theorem 2, respectively.            
\begin{corollary}[Maximum Diversity Estimates]\label{C3}
The maximum secrecy diversity estimates provided by $\hat{d}_{s}^{U}$ and $\hat{d}_{s}^{L}$ are respectively given by:
\begin{IEEEeqnarray}{rCl}
\hat{d}_{s}^{U,max}&=&\underset{r_{s}\rightarrow 0} \lim{d_{s}^{U}(r_{s})} \nonumber \\
                   &=& m \  k \left(1-\frac{m-1}{2k}\right)\left(1-\frac{g \eta}{\left(1+g\eta\right)\ln{\left(1+g\eta\right)}}\right) \label{E51} \\
\hat{d}_{s}^{L,max}&=&\underset{r_{s}\rightarrow 0} \lim{d_{s}^{L}(r_{s})} \nonumber \\
                   &=& m \  k \left(1-\frac{g \eta}{\left(1+g\eta\right)\ln{\left(1+g\eta\right)}}\right) \label{E52}
\end{IEEEeqnarray}
\end{corollary}
\begin{IEEEproof}
For convenience, the proof is presented in Appendix \ref{App1}
\end{IEEEproof}
By letting $\eta$ tends toward zero in (\ref{E51}) and (\ref{E52}), it can be seen that $\hat{d}_{s}^{U,max}$ underestimates the maximum diversity gain at high-SNR, which is equal to $m k$; whereas, $\hat{d}_{s}^{L,max}$ matches the later asymptotic maximum diversity gain. In fact, at high-SNR regime, $\hat{d}_{s}^{U}$ underestimates the diversity gain at any multiplexing gain, whereas, $\hat{d}_{s}^{L}$ matches the asymptotic secrecy diversity gain at all multiplexing gains, as stated below.
\begin{theorem}[High-SNR Estimates of Secrecy Diversity Gain]\label{T3}
At high-SNR, we have:
\begin{IEEEeqnarray}{rCl}
\underset{\eta \rightarrow \infty} \lim{d_{s}^{U}\left(r_{s}\right)} &=& 
\begin{cases}
 m \ k \left(1-\frac{m-1}{2k}\right)-\left[m \ k \left(1-\frac{m-1}{2k}\right)-\frac{m-1}{\overset{m}{\underset{l=1}\sum}{\frac{1}{k-l+1}}}\right]  r_{s}  \\
 \quad \quad \quad \quad \quad \quad \quad \quad  \quad \quad \quad   \text{if} \quad r_{s} \in [0,1[ \\
 \frac{1}{\overset{m}{\underset{l=1}\sum}\frac{1}{k-l+1}}\left(m-r_{s}\right)       \quad \quad \quad   \quad \quad \text{if} \quad r_{s} \in [1,m]
 \end{cases} \label{E53}
 \\
\underset{\eta \rightarrow \infty} \lim{d_{s}^{L}(r_{s})} &=&  d_{s}^{\infty}\left(r_{s}\right) \label{E54}
\end{IEEEeqnarray}
\end{theorem}
\begin{IEEEproof}
For convenience, the proof is presented in Appendix \ref{App2}
\end{IEEEproof}

\begin{figure*}[!t]
\setcounter{mytempeqncnt}{\value{equation}}
\setcounter{equation}{27}
\small
\begin{IEEEeqnarray}{rCl}
     \mu_{eq} &=&  \sum_{n=0}^{m-1}\frac{e^{1/\eta}(k-m+n)!}{n!(k- m)!^2} \sum_{l=0}^{2n}\frac{(k-m)!\Gamma(l-n)}{\Gamma(-n)l!} \nonumber \\
              &\quad&     \times  \sum_{q=1}^{k-m+l+1}\eta^{-1 - l + m - k + q} \Gamma\Bigl(-1 - l + m - k + q, 1/\eta\Bigr)   \times  _{3}F_{2}\Bigl(-n, -l, -l + m - k;1 + n - l, 1 - m + k; -1\Bigr) \label{E61} \\
     \sigma_{eq}^{2} &=& \sum _{n=0}^{m-1} \frac{e^{1/\eta } (k-m+n)!}{n! (k-m)!^2} \sum _{l=0}^{2 n} \left(\frac{2  (k-m)!\Gamma(l-n) }{ (k-m+l)! l!\Gamma(-n)} \times \, _3F_2(-n,-l,-l+m-k;n-l+1,-m+k+1;-1)  \right. \nonumber \\
            & \quad& \left. \times \sum _{q=0}^{l-m+k} (-1)^q \left(
\begin{array}{c}
 l-m+k \\
 q
\end{array}
\right) G_{3,4}^{4,0}\left(\frac{1}{\eta }\Bigg|
\begin{array}{c}
 q+1,q+1,q+1 \\
 l-m+k+1,q,q,q
\end{array}
\right) \right) \nonumber \\
             &\quad&  -\sum _{i=0}^{m-1} \left(\sum _{j=0}^{m-1} \left(\frac{e^{2/\eta } (k-m+i)! (k-m+j)!}{i! j! (k-m+1)!^4}  \left(\sum _{n=0}^i \left(\sum _{l=0}^j \left(\frac{  (k-m)!^2 (k-m+n+l)! \Gamma(n-i) \Gamma(l-j)}{ n! l! (k-m+n)! (k-m+l)! \Gamma(-i) \Gamma(-j)} \times \sum _{q=1}^{n+l-m+k+1} E_{n+l-m+k-q+2}\left(\frac{1}{\eta }\right)  \right) \right)             \right)^2 \right)\right) \label{E62}       
\end{IEEEeqnarray}
\normalsize
\setcounter{equation}{\value{mytempeqncnt}}
\hrulefill
\end{figure*}

\section{ZF Outage Analysis Using A Gaussian Approximation}\label{S6}
Instead of the bounding techniques used in section \ref{S5}, we present in this section performance of ZF transmit scheme using a Gaussian approximation. Gaussian approximation of the channel mutual information $\Psi{\left(\textbf{H}_{eq}\right)}$, initially proposed by Smith and Shafi in \cite{Smith}, has been widely used since then, albeit in a different context, to provide a more tractable analytical tool. While this approximation does not provide much insight as to how secure DMT of ZF transmit scheme varies with SNR and with the multiplexing gain $r_{s}$, the accuracy ot the results as shown by simulation (cf. section \ref{S7}), especially at finite-SNR, and the analytical tractability, both justify its presentation in this section. 
\subsection{Outage Probability Approximation}\label{S61}
Recall that the mean $\mu_{eq}$ and the variance $\sigma_{eq}^{2}$ of $\Psi{\left(\textbf{H}_{eq}\right)}$ have been derived in integral forms by Telatar in \cite{Telatar} and by Smith and Shafi in \cite{Smith}, respectively. Kang and Alouini succeeded to obtain these quantities in closed forms \cite[eqs. (34) and (35)]{Kang}. Introducing the generalized hypergeometric function $ _{p}F_{q}(a;b;z)$, the Meijer's G function $G_{p,q}^{m,n}\left(z \Bigg|
\begin{array}{c}
 a_1,\ldots,a_p \\
 b_1,\ldots,b_q
\end{array}
\right) $, and the exponential integral function $E_{n}(z)$ \cite{Ryzhik}, it can be shown that their expressions may be further simplified as shown at the top of the next page, where the Gamma function in (\ref{E61}) is defined by: $\Gamma\left(a, z\right)=\int_{z}^{\infty}t^{a-1}e^{-t}\ dt$. Using the above Gaussian approximation, the upper bound in (\ref{E33}) may be approximated by:
\addtocounter{equation}{2}
\begin{equation}\label{E63}
P_{out}^{ZF} \approx \Q{ \Biggl(\frac{\mu_{eq}-R_{s}}{\sqrt{\sigma_{eq}^{2}}}\Biggr)},
\end{equation}
where $\Q{(\cdot)}$ represents the Gaussian Q-function.

\subsection{Secure DMT Approximation}\label{S62}
Using (\ref{E63}) along with (\ref{E27}), an estimate of ZF secure diversity gain $\hat{d}_{s}^{G}(r_{s},\eta)$ (G for Gaussian) can be expressed by:
\begin{equation}\label{E64}	
\hat{d}_{s}^{G}(r_{s},\eta)=\frac{\eta}{ \Q{ \Biggl(\frac{\mu_{eq}-R_{s}}{\sqrt{\sigma_{eq}^{2}}}\Biggr)}}\ \frac{1}{\sqrt{2\pi}}\ e^{-\frac{(R_{s}-\mu_{eq})^{2}}{2\sigma_{eq}^{2}}}h(r_{s},\eta),
\end{equation}
where $h(r_{s},\eta)$ is defined by:
\begin{equation*}
h(r_{s},\eta)=\frac{(R_{s}-\mu_{eq})(\sigma_{eq}^{2})^{\prime}}{2(\sigma_{eq}^{2})^{3/2}}-\frac{R_{s}^{\prime}-\mu_{eq}^{\prime}}{(\sigma_{eq}^{2})^{1/2}},
\end{equation*}
where $f^{\prime}$ represents the derivative of $f$ with respect to $\eta$. Expressions of $R_{s}^{\prime}$, $\mu_{eq}^{\prime}$ and $(\sigma_{eq}^{2})^{\prime}$ are given by (\ref{E65}), (\ref{E66}) and (\ref{E67}), respectively, in Appendix \ref{App3}.\ 

Using the asymptotic representation of the Q-function: $\Q(x)=e^{-x^2/2}\left(\frac{1}{2 \sqrt{\pi}x}+o\left(\frac{1}{x}\right)\right)$, letting $t=\frac{\mu_{eq}-R_{s}}{\sqrt{\sigma_{eq}^{2}}}$ for convenience, and applying the $\log{(\cdot)}$ function to the RHS of (\ref{E63}), we have at high-SNR:
\begin{equation}\label{E68} 
\log{\left(\Q{ \left(t\right)} \right)}= -\frac{t^2}{2}+o(t). 
\end{equation}
But at high-SNR, $t\dot{=}\left(m-r_{s}\right)\log{\left(\eta\right)}$,\footnote{The notation $f\dot{=}g $ means $\underset{SNR\rightarrow \infty}{\lim}{\frac{\log{f}}{\log{SNR}}}=\underset{SNR\rightarrow \infty}{\lim}{\frac{\log{g}}{\log{SNR}}}$} which combined with (\ref{E68}) give:
\begin{equation}\label{E69} 
\log{\left(\Q{ \left(t\right)} \right)}\dot{=} -\frac{\left(r_{s}-m\right)^2}{2}\log{\left(\eta\right)}^2. 
\end{equation} 
That is, the asymptotic secrecy diversity gain estimate provided by the Gaussian approximation diverges for all multiplexing gains.

\section{Numerical results}\label{S7}
\begin{figure}[t]
  \begin{center}
    \includegraphics[scale=0.27]{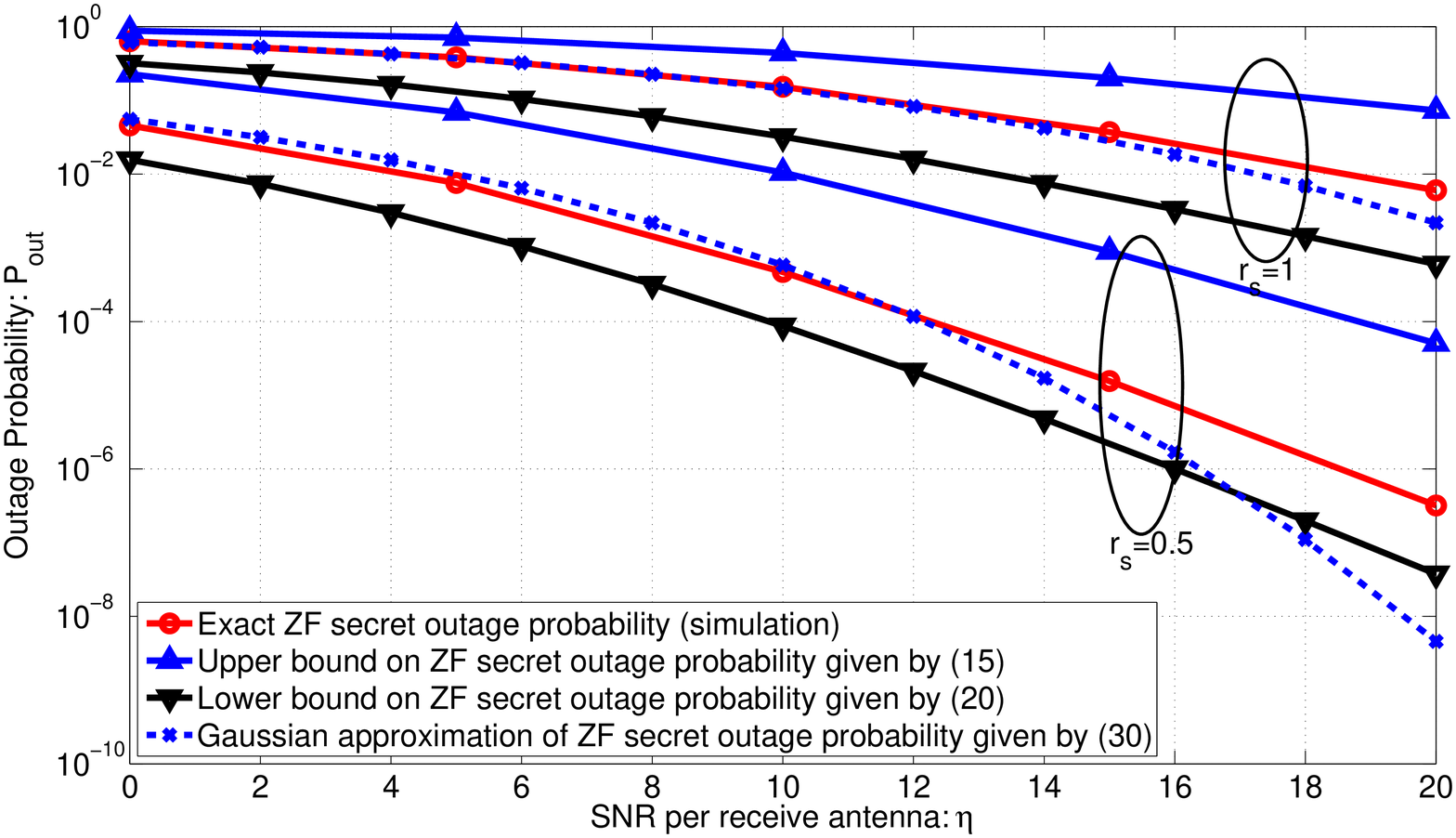}
    \caption{Comparison of ZF secrecy outage probability, the upper bound given by (\ref{E41}), the lower bound given by (\ref{E46}) and the Gaussian approximation given by (\ref{E63}); for different secrecy multiplexing gains $r_{s}$ and for a wiretap channel with $\left(N_{t},N_{m},N_{e}\right)=\left(4,2,1\right)$.}
    \label{F1}
  \end{center}
\end{figure}

\begin{figure}[t]
  \begin{center}
    \includegraphics[scale=0.27]{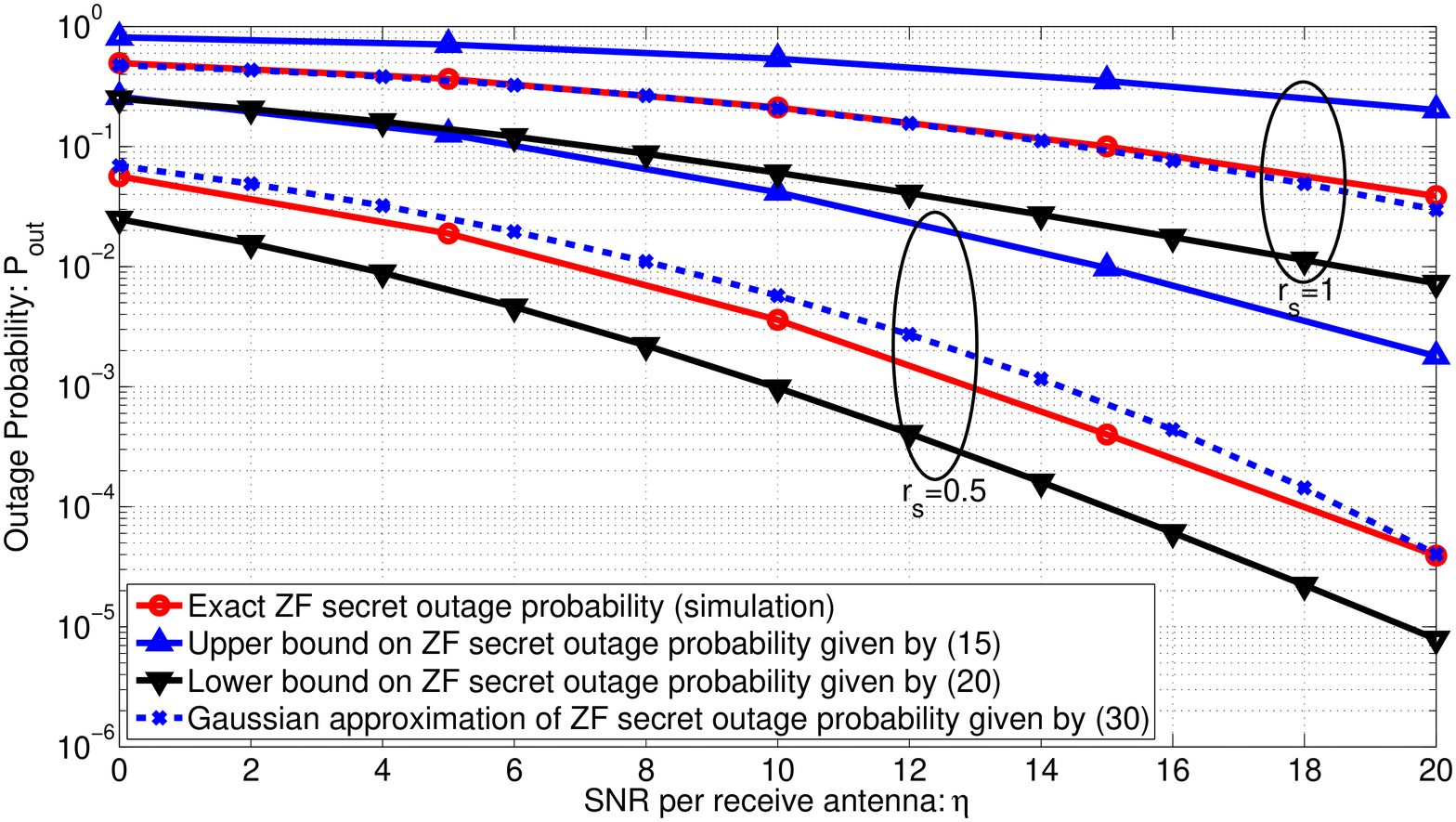}
    \caption{Comparison of ZF secrecy outage probability, the upper bound given by (\ref{E41}), the lower bound given by (\ref{E46}) and the Gaussian approximation given by (\ref{E63}); for different secrecy multiplexing gains $r_{s}$ and for a wiretap channel with $\left(N_{t},N_{m},N_{e}\right)=\left(3,2,1\right)$.}
    \label{F2}
  \end{center}
\end{figure}

\begin{figure}[t]
  \begin{center}
    \includegraphics[scale=0.27]{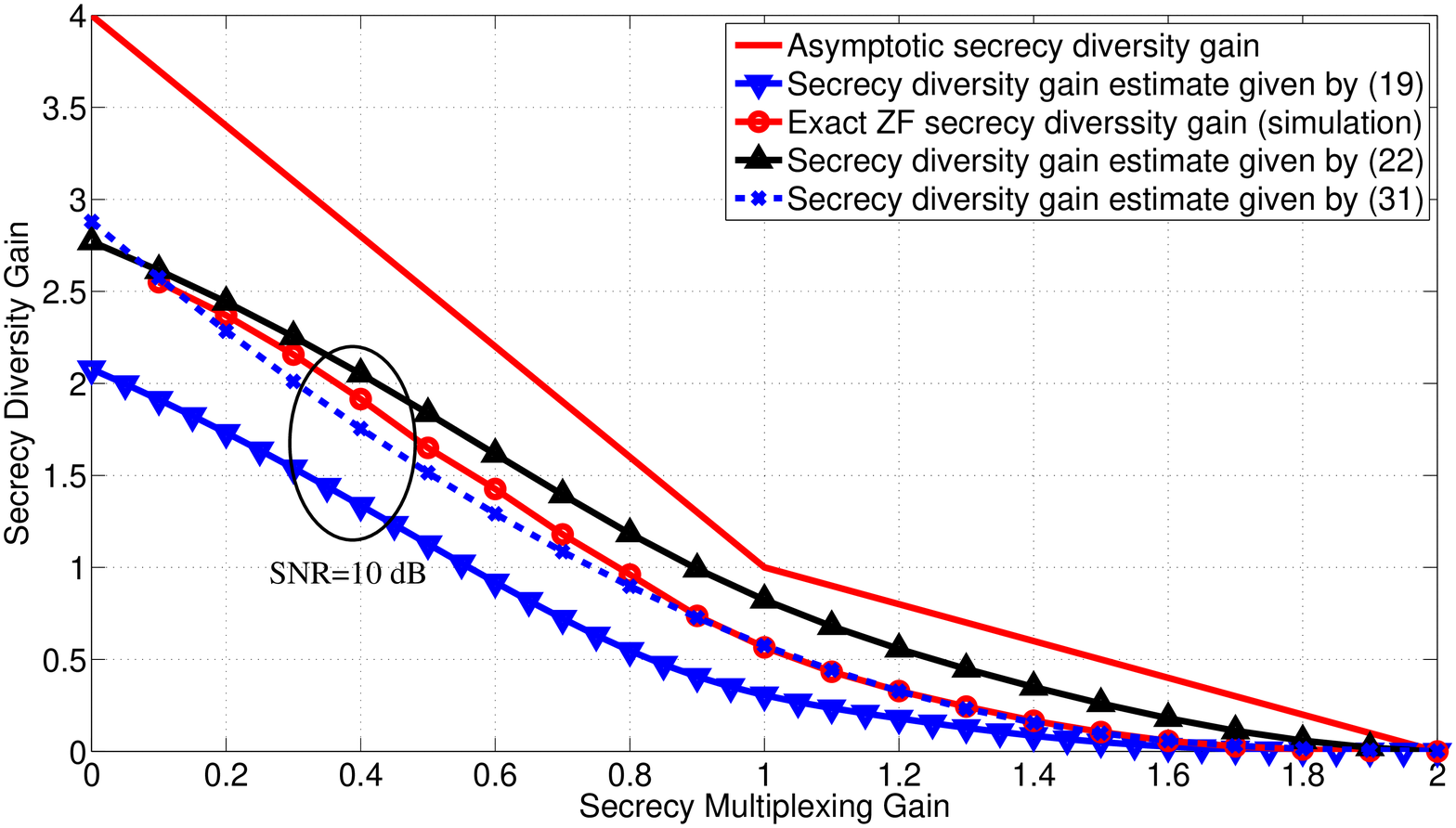}
    \caption{Comparison of ZF secrecy diversity gain obtained by simulation and its estimates $\hat{d}_{s}^{U}\left(r_{s},\eta\right)$, $\hat{d}_{s}^{L}\left(r_{s},\eta\right)$ and $\hat{d}_{s}^{G}(r_{s},\eta)$, given by (\ref{E45}), (\ref{E48}) and (\ref{E63}), respectively; for an SNR value $\eta=10$ dB, and for a wiretap channel with $\left(N_{t},N_{m},N_{e}\right)=\left(3,2,1\right)$.}
    \label{F3}
  \end{center}
\end{figure}

\begin{figure}[t]
  \begin{center}
    \includegraphics[scale=0.27]{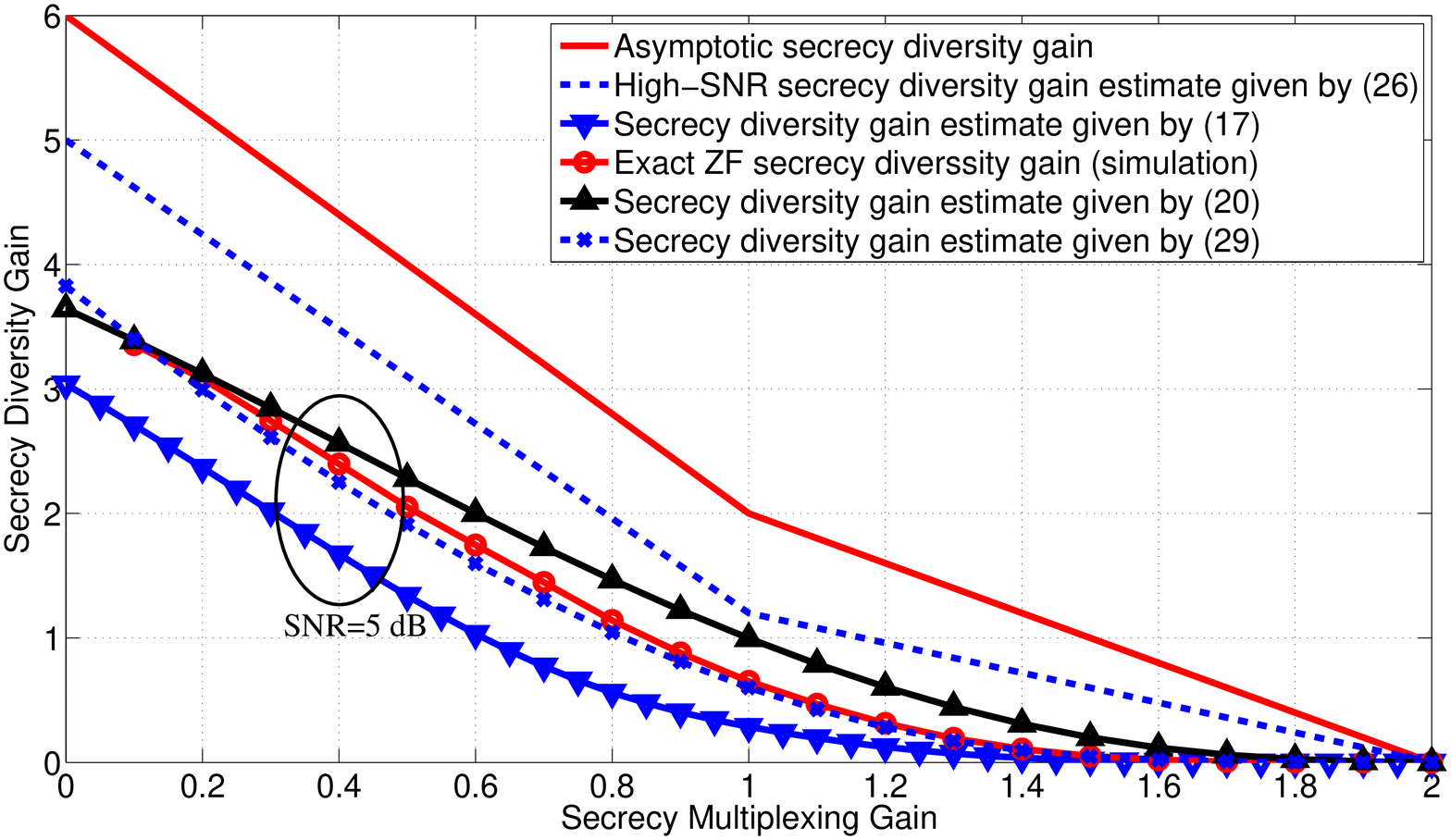}
    \caption{Comparison of ZF secrecy diversity gain obtained by simulation and its estimates $\hat{d}_{s}^{U}\left(r_{s},\eta\right)$, $\hat{d}_{s}^{L}\left(r_{s},\eta\right)$ and $\hat{d}_{s}^{G}(r_{s},\eta)$, given by (\ref{E45}), (\ref{E48}) and (\ref{E63}), respectively; for an SNR value $\eta=5$ dB, and for a wiretap channel with $\left(N_{t},N_{m},N_{e}\right)=\left(4,2,1\right)$.}
    \label{F4}
  \end{center}
\end{figure}

\begin{figure}[t]
  \begin{center}
    \includegraphics[scale=0.27]{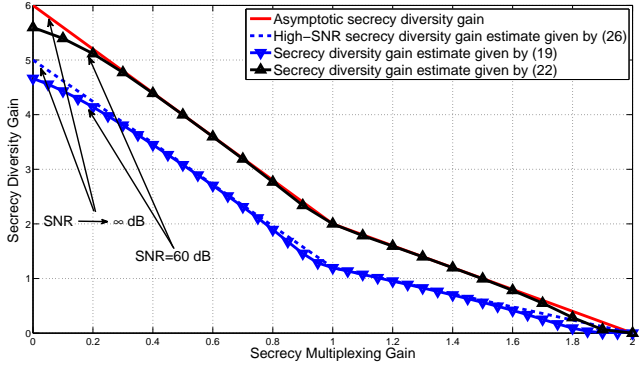}
    \caption{Behavior of ZF diversity estimates $\hat{d}_{s}^{U}\left(r_{s},\eta\right)$ and $\hat{d}_{s}^{L}\left(r_{s},\eta\right)$, for an  SNR value $\eta=60$ dB and for a wiretap channel with $\left(N_{t},N_{m},N_{e}\right)=\left(4,2,1\right)$.}
    \label{F5}
  \end{center}
\end{figure}

In this section, numerical results of ZF transmit scheme for a MIMO wiretap channel with $\left(N_{t},N_{m},N_{e}\right)=\left(3,2,1\right)$ and $\left(N_{t},N_{m},N_{e}\right)=\left(4,2,1\right)$, are presented. In Fig. \ref{F1} and Fig. \ref{F2}, ZF secrecy outage probability obtained using computer simulation, along with the upper, the lower bounds and the Gaussian approximation given respectively by (\ref{E41}), (\ref{E46}) and (\ref{E63}), are plotted for secrecy multiplexing gains $r_{s}=0.5$ and $r_{s}=1$ and for $\left(N_{t},N_{m},N_{e}\right)=\left(3,2,1\right)$ and $\left(N_{t},N_{m},N_{e}\right)=\left(4,2,1\right)$, respectively. As can be seen in Fig. \ref{F1} and Fig. \ref{F2}, the curves corresponding to the bounds follow the same shape as the exact curves for all SNR values, supporting the results reported in Theorem \ref{T1} and Theorem \ref{T2}. The lower bound curves are however closer to the exact curves than the upper bound ones, as shown in Fig. \ref{F1} and Fig. \ref{F2}. The Gaussian approximation curves are enough accurate to approximate the ZF secrecy outage probability at low SNR (below 12 dB in Fig. \ref{F1} and up to 20 dB in Fig. \ref{F2}). However, as SNR increases, a discrepancy between the Gaussian approximation curves and the curves obtained by simulations shows up at high-SNR, suggesting that the Gaussian approximation is somehow too pessimistic to be useful in this case. The secrecy diversity gain and its estimates given by (\ref{E45}), (\ref{E48}) and (\ref{E64}) are reported in Fig. \ref{F3} and Fig. \ref{F4} for different antenna configurations and different SNR values. In these figures, it can be seen that the asymptotic secrecy diversity gain provides an optimistic upper bound on the secrecy diversity gain at finite-SNR. Furthermore, the secrecy diversity estimates provided by the upper bound is closer to the exact secrecy diversity gain at high multiplexing gains ($r_{s}\geq 1$). As the multiplexing gain decreases, the secrecy diversity gain estimate provided by the lower bound is more accurate and converges to the exact curves as $r_{s}\rightarrow 0$. In both Fig. \ref{F3} and Fig. \ref{F4}, the secrecy diversity gain estimate provided by the Gaussian approximation is the closest to the exact curves for almost all multiplexing gains which again justify the presentation of this approximation in this work. Also shown in Fig. \ref{F4}, is the asymptotic (at high-SNR) diversity estimate $\hat{d}_{s}^{U}$. Finally, the asymptotic behavior of the diversity estimates is illustrated in Fig. \ref{F5}, where the diversity estimate curves are plotted for an SNR $\eta=60$ dB. Figure \ref{F5} confirms the results presented in Theorem \ref{T3}.

\section{Conclusion}\label{S8}
In this paper, the finite-SNR DMT of ZF transmit scheme under secrecy constraint has been analyzed. Performance limits of such a transmit scheme have been characterized in terms of secrecy outage probability, where analytically tractable upper and lower bounds have been derived. Related secrecy diversity gain estimates have been provided. These diversity estimates give insight as to how the performance of ZF varies with SNR and with the secrecy multiplexing gain. Through asymptotic analysis, we characterize ZF transmit scheme in some limit conditions. Motivated by its accuracy, we have also presented performance of the Gaussian approximation and its related diversity estimates. Simulation results have been provided to confirm the accuracy of the finite-SNR DMT characterizations. Furthermore, beside the fact that the diversity estimates can be used to characterize the potential limit of a particular transmit scheme, it can also be used, for instance, as a metric to improve error performance of certain adaptive communications under secrecy constraint.

\appendices
\section{Proof of Corollary \ref{C3}}\label{App1}
First, note that since $r_{s}$ tends toward zero, then so are all $b_{l}$, $l=1,\ldots,m$. Then, it is easy to verify that:
\begin{equation}\label{EA11}
\underset{r_{s}\rightarrow 0} \lim \ {\frac{\beta_{l}}{\alpha_{l}}\biggl(f_{l}(b_{l})-f_{l}(r_{s})\biggr)\frac{\underset{k\neq l}\prod\left(1-\frac{\beta_{k}}{\alpha_{k}}\right)}{1-\overset{m} {\underset{k=1} \prod}\left(1-\frac{\beta_{k}}{\alpha_{k}}\right)}}=0.
\end{equation}
This follows from the facts that:
\begin{equation}\label{EA12}
0 \leq \frac{\beta_{l}}{\alpha_{l}}\frac{\underset{k\neq l}\prod\left(1-\frac{\beta_{k}}{\alpha_{k}}\right)}{1-\overset{m} {\underset{k=1} \prod}\left(1-\frac{\beta_{k}}{\alpha_{k}}\right)} \leq 1,
\end{equation}
along with: 
\begin{IEEEeqnarray}{rcl}
\underset{b_{l}\rightarrow 0} \lim \ {f_{l}(b_{l})}&=&\left(1-\frac{g\eta}{\left(1+g\eta\right)\log{\left(1+g\eta\right)}}\right)\left(k-l+1\right) \label{EA13} \\
                                                &=& \underset{r_{s}\rightarrow 0} \lim \ {f_{l}(r_{s})}. \label{EA14}
\end{IEEEeqnarray}
Equation (\ref{EA11}) implies that this term does not provide any diversity contribution at low secrecy multiplexing gain. Using (\ref{EA11}) and (\ref{EA14}), (\ref{E51}) is immediately obtained. To obtain (\ref{E52}), a similar approach can be used.

\begin{figure*}[!t]
\setcounter{mytempeqncnt}{\value{equation}}
\setcounter{equation}{50}
\begin{alignat}{3}
R^{\prime}&=\frac{t N_{e}}{1+N_{e} \eta }+ \frac{r_{s} N_{m}}{1+N_{m} \eta }\label{E65}\\
\mu_{eq}^{\prime} &=\sum _{k=0}^{m-1} \frac{e^{1/\eta } (n-m+k)!}{k! (n-m)!^2} \sum _{l=0}^{2 k} \frac{\Gamma (l-k) (n-m)!}{\Gamma (-k) l!} \nonumber \\
                 & \quad \times \sum _{q=1}^{n-m+l+1} \left((-l+m-n+q-1) \eta ^{-l+m-n+q-2} \Gamma \left(-l+m-n+q-1,\frac{1}{\eta }\right)+\frac{1}{e^{1/\eta } \eta }\right) \nonumber \\
                 & \quad \times \, _3F_2(-k,-l,-l+m-n;k-l+1,-m+n+1;-1) \nonumber \\
                 & -\frac{e^{1/\eta } (n-m+k)!}{\eta ^2 k! (n-m)!^2} \sum _{l=0}^{2 k} \frac{\Gamma (l-k) (n-m)!}{\Gamma (-k) l!} \sum _{q=1}^{l-m+n+1} \eta ^{-l+m-n+q-1} \Gamma \left(-l+m-n+q-1,\frac{1}{\eta }\right) \nonumber \\
                 & \quad \times \, _3F_2(-k,-l,-l+m-n;k-l+1,-m+n+1;-1) \label{E66}\\
(\sigma_{eq}^{2})^{\prime} &= \sum _{k=0}^{m-1} \left(\frac{e^{1/\eta } (n-m+k)!}{(n-m)!^2k!} \sum _{l=0}^{2 k} \frac{2 \Gamma (l-k)  (n-m)!}{\Gamma (-k) (n-m+l)!l!} \right.\nonumber\\
         & \quad\left. \times \, _3F_2(-k,-l,-l+m-n;k-l+1,-m+n+1;-1) \right.\nonumber \\
         & \quad\left. \times \sum _{q=0}^{l-m+n} -\frac{(-1)^q }{\eta }  \left(q G_{3,4}^{4,0}\left(\frac{1}{\eta }\Bigg|
\begin{array}{c}
 q+1,q+1,q+1 \\
 l-m+n+1,q,q,q
\end{array}
\right)-G_{2,3}^{3,0}\left(\frac{1}{\eta }\Bigg|
\begin{array}{c}
 q+1,q+1 \\
 l-m+n+1,q,q
\end{array}
\right)\right)\right.\nonumber \\
        & \quad\left. -\frac{e^{1/\eta } (n-m+k)!}{\eta ^2  (n-m)!^2k!} \sum _{l=0}^{2 k} \frac{2 \Gamma (l-k) (n-m)!}{\Gamma (-k) (n-m+l)!l!}\right.\nonumber \\
         & \quad\left. \times \, _3F_2(-k,-l,-l+m-n;k-l+1,-m+n+1;-1)\right.\nonumber\\
         & \quad\left. \times \sum _{q=0}^{l-m+n} (-1)^q \left(
\begin{array}{c}
 l-m+n \\
 q
\end{array}
\right) G_{3,4}^{4,0}\left(\frac{1}{\eta }\Bigg|
\begin{array}{c}
 q+1,q+1,q+1 \\
 l-m+n+1,q,q,q
\end{array}
\right)\right)\nonumber \\
        & \quad -\sum _{i=0}^{m-1} \left(\sum _{j=0}^{m-1} \frac{2 e^{2/\eta } (n-m+i)! (n-m+j)!}{\eta ^2  (n-m)!^4i!j!}\right.\nonumber\\
        & \quad\left. \times \left(\sum _{k=0}^i \left(\sum _{l=0}^j \frac{\Gamma (k-i) \Gamma (l-j) (n-m)!^2 (n-m+k+l)!}{\Gamma (-i) \Gamma (-j) \Gamma (k+1) \Gamma (l+1) \Gamma (k-m+n+1) \Gamma (l-m+n+1)} \sum _{q=1}^{k+l-m+n+1} E_{k+l-m+n-q+2}\left(\frac{1}{\eta }\right)\right)\right)\right.\nonumber \\
        & \quad\left. \times \left( \eta ^2 \sum _{k=0}^i \left(\sum _{l=0}^j \frac{\Gamma (k-i) \Gamma (l-j) (n-m)!^2 (n-m+k+l)!}{\Gamma (-i) \Gamma (-j) (n-m+k)! (n-m+l)!k! l! }\right.\right.\right.\nonumber \\
        & \quad\left.\left.\left. \times  \sum _{q=1}^{k+l-m+n+1} \frac{e^{-1/\eta }-(k+l-m+n-q+1) E_{k+l-m+n-q+2}\left(\frac{1}{\eta }\right)}{\eta }\right)\right.\right.\nonumber \\
        & \quad\left.\left. -\sum _{k=0}^i \left(\sum _{l=0}^j \frac{\Gamma (k-i) \Gamma (l-j) (n-m)!^2 (n-m+k+l)!}{\Gamma (-i) \Gamma (-j)  (n-m+k)! (n-m+l)! k! l!} \sum _{q=1}^{k+l-m+n+1} E_{k+l-m+n-q+2}\left(\frac{1}{\eta }\right)\right)\right)\right) \label{E67}
\end{alignat}
\setcounter{equation}{\value{mytempeqncnt}}
\hrulefill
\end{figure*}

\section{Proof of Theorem \ref{T3}}\label{App2}
Theorem \ref{T3} can be proved in two different ways. The first one analyzes the expressions of the upper and the lower bounds given by (\ref{E45}) and (\ref{E48}), respectively, at high-SNR regime. The second one achieves the same goal by deriving the asymptotic behavior of the diversity estimates in Corollary \ref{C1} and Corollary \ref{C2}, respectively, as SNR tends toward infinity. We will present these two approaches to prove (\ref{E53}), the proof of (\ref{E54}) follows along similar lines.
\begin{enumerate}
\item Method 1: Analyzing the upper bound (\ref{E45}) at high-SNR
\end{enumerate}
We distinguish two cases: $r_{s} <1$ and $r_{s} \geq 1$.
\begin{itemize}
\item $r_{s} <1$
\end{itemize}  
First, note that as $\eta \rightarrow \infty$, $\xi\left(r_{s}\right) \rightarrow 0$ and so is $\xi\left(b_{l}\right)$, $l=1,\dots,m$. 
On the other hand, since around zero, $\Gamma_{inc}\left(\cdot,\cdot\right)$ may be approximated by: $\Gamma_{inc}\left(x,N\right)=\frac{x^N}{N!}+ o\left(x^N\right)$, then we have:
\begin{equation}\label{EA21}
\alpha_{l} = \frac{1}{\left(k-l+1\right)!}\left(\left(N_{t}-N_{e}\right)g^{r_{s}}\eta^{r_{s}-1}\right)^{k-l+1} +  o\left(\eta^{\left(r_{s}-1\right)\left(k-l+1\right)}\right) 
\end{equation}
\begin{equation}\label{EA22 }
\beta_{l} = \frac{1}{\left(k-l+1\right)!}\left(\left(N_{t}-N_{e}\right)g^{b_{l}}\eta^{b_{l}-1}\right)^{k-l+1} +  o\left(\eta^{\left(b_{l}-1\right)\left(k-l+1\right)}\right)
\end{equation}
\begin{equation}\label{EA23}  
\frac{\beta_{l}}{\alpha_{l}}=\left(g\eta\right)^{\left(b_{l}-r_{s}\right)\left(k-l+1\right)}+o\left(\eta^{\left(b_{l}-r_{s}\right)\left(k-l+1\right)}\right)
\end{equation}
\begin{equation}\label{EA24}
  \left(1-\prod_{l=1}^{m}\Bigl[1-\frac{\beta_{l}}{\alpha_{l}}\Bigr]\right) = \sum_{l=1}^{m}\left(g\eta\right)^{\left(b_{l}-r_{s}\right)\left(k-l+1\right)}+o\left(\eta^{\underset{b_{l}}\max\{\left(b_{l}-r_{s}\right)\left(k-l+1\right)\}}\right)  
\end{equation}
Moreover, the first term $\left(\prod_{l=1}^{m}\alpha_{l}\right)$ on the right  hand side (RHS) of (\ref{E41}) does not depend on $b_{l}$'s, and from (\ref{EA21}), provides an asymptotic diversity gain equal to: $\left(1-r_{s}\right)\sum_{l=1}^{m}\left(k-l+1\right)$. Consequently, at high-SNR, minimizing (\ref{E41}) is equivalent to minimizing RHS of (\ref{EA24}). The problem at hand now reduces to solving the following minimization problem:
\begin{equation}\label{EA25}
\begin{cases}
\underset{b_{l}}\min{ \ \sum_{l=1}^{m}\eta^{\left(b_{l}-r_{s}\right)\left(k-l+1\right)}} \\
0 \geq  b_{1}\geq \ldots \geq b_{m} \\
 \sum_{l=1}^{m}{b_{l}}=r_{s}
\end{cases}
\end{equation}
This problem is convex and can be easily solved by the Lagrangian method. We skip the details and give below the optimal solution at high-SNR:
\begin{equation}\label{EA26}
\left(b_{l}-r_{s}\right)\left(k-l+1\right)= - \delta,
\end{equation}
where $\delta = \frac{m-1}{\overset{m}{\underset{l=1}\sum}{\frac{1}{k-l+1}}}r_{s}$, for all $b_{l}$, $l=1,\ldots,m$. Thus, the asymptotic diversity gain provided by the second term on the RHS of (\ref{E41}) is equal to $\delta$. Summing the two diversity gains, we have:
\begin{equation}\label{EA28}
 \underset{\eta \rightarrow \infty} \lim{d_{s}^{U}\left(r_{s}\right)}=\left(1-r_{s}\right)\sum_{l=1}^{m}\left(k-l+1\right) + \frac{m-1}{\overset{m}{\underset{l=1}\sum}{\frac{1}{k-l+1}}}r_{s},
\end{equation}
for $r_{s} \in [0,1[$, from which the first relation in (\ref{E53}) is immediate.
\begin{itemize}
\item $r_{s} \geq 1$
\end{itemize} 
On the contrary to the first case, when $\eta \rightarrow \infty$, $\alpha_{l}$ tends toward 1, $l=1,\ldots,m$, which implies that the first term $\left(\prod_{l=1}^{m}\alpha_{l}\right)$ on the right  hand side (RHS) of (\ref{E41}) does not provide any diversity gain in this case. Hence, minimizing the upper bound (\ref{E41}) is equivalent to solving the following optimization problem at high-SNR:  
\begin{equation}\label{EA29}
\begin{cases}
\underset{b_{l}}\min{ \ \sum_{l=1}^{m}\eta^{\left(b_{l}-1\right)\left(k-l+1\right)}} \\
0 \geq  b_{1}\geq \ldots \geq b_{m} \\
 \sum_{l=1}^{m}{b_{l}}=r_{s},
\end{cases}
\end{equation}
which can be solved similarly to the first case to obtain the following optimal solution:
\begin{equation}\label{EA30}
\left(b_{l}-1\right)\left(k-l+1\right)= - \gamma,
\end{equation}
where $\gamma = \frac{m-r_{s}}{\overset{m}{\underset{l=1}\sum}{\frac{1}{k-l+1}}}$ for all $b_{l}$, $l=1,\ldots,m$. Hence, for $r_{s} \in [1,m]$, the asymptotic secrecy diversity gain is equal to $\gamma$, which proves the second relation in (\ref{E53}) too.
\begin{enumerate}
\item Method 2: Analyzing the diversity estimates in Corollary \ref{C1} at high-SNR
\end{enumerate}
We also treat the cases $r_{s} <1$ and $r_{s} \geq 1$ separately.
\begin{itemize}
\item $r_{s} <1$
\end{itemize}  
First, it can be easily verified that:
\begin{IEEEeqnarray}{rcl}
f_{l}(r_{s})&=&\frac{\eta}{N_{t}-N_{e}}\left(k-l+1\right)\left(1-r_{s}\right) + o\left(\eta\right) \label{EA32} \\
f_{l}(b_{l})&=&\frac{\eta}{N_{t}-N_{e}}\left(k-l+1\right)\left(1-b_{l}\right)+ o\left(\eta\right). \label{EA33}
\end{IEEEeqnarray}
Thus the diversity contribution of the first term $\frac{N_{t}-N_{e}}{\eta}\sum_{l=1}^{m} f_{l}(r_{s})$ on the RHS of (\ref{E45}) is equal to $\left(1-r_{s}\right)\sum_{l=1}^{m}\left(k-l+1\right)$. We are then left to analyze the second term on the RHS of (\ref{E45}). Combining (\ref{EA23}), (\ref{EA24}), (\ref{EA32}) and (\ref{EA33}), along with tha fact that $\prod_{k\neq l}\left(1-\frac{\beta_{l}}{\alpha_{l}}\right) \rightarrow 1$, as SNR tends toward infinity, the limit as $ \eta \rightarrow \infty$, of the diversity estimate in (\ref{E45}) can be expressed by:
\begin{IEEEeqnarray}{rcl}
\underset{\eta \rightarrow \infty} \lim{d_{s}^{U}\left(r_{s}\right)}&=&\left(1-r_{s}\right)\sum_{l=1}^{m}\left(k-l+1\right) \nonumber \\            
&\quad& +\frac{\sum_{l=1}^{m}\left(r_{s}-b_{l}\right)\left(k-l+1\right)\eta^{\left(b_{l}-r_{s}\right)\left(k-l+1\right)}}{\sum_{l=1}^{m}\eta^{\left(b_{l}-r_{s}\right)\left(k-l+1\right)}} +o(\eta) \label{EA34}.
\end{IEEEeqnarray}
Recall that $b_{l}$'s in (\ref{EA34}) are those that solve (\ref{EA25}), and thus by (\ref{EA26}), we have:
\begin{equation}\label{EA35}
\underset{\eta \rightarrow \infty} \lim{d_{s}^{U}\left(r_{s}\right)}=\left(1-r_{s}\right)\sum_{l=1}^{m}\left(k-l+1\right) +\delta,
\end{equation} 
from which the first relation in (\ref{E53}) follows.
\begin{itemize}
\item $r_{s} \geq 1$
\end{itemize}  
To prove (\ref{E54}), we use the fact that when $\eta \rightarrow \infty$, $\alpha_{l}$ tends toward 1, $l=1,\ldots,m$, along with the fact that $f_{l}\left(r_{s}\right)=o\left(\eta^{-N}\right)$, for all $N \geq 0$. The proof is then completed using similar machinery than the one utilized in the first case.

\section{Expressions of $R_{s}^{\prime}$, $\mu_{eq}^{\prime}$ and $(\sigma_{eq}^{2})^{\prime}$}\label{App3}
Expressions of $R_{s}^{\prime}$, $\mu_{eq}^{\prime}$ and $(\sigma_{eq}^{2})^{\prime}$ are given by (\ref{E65}), (\ref{E66}) and (\ref{E67}), respectively, at the top of the page.

\begin{biography}{Zouheir Rezki} (S'01, M'08) was born in
Casablanca, Morocco. He received the Diplome
d'Ing\'enieur degree from the \'Ecole Nationale de
l'Industrie Min\'erale (ENIM), Rabat, Morocco, in
1994, the M.Eng. degree from  \'Ecole de Technologie
Sup\'erieure, Montreal, Qu\'ebec, Canada, in 2003, and
the Ph.D. degree from  \'Ecole Polytechnique, Montreal,
Qu\'ebec, in 2008, all in electrical engineering.
From October 2008 to September 2009, he was a
postdoctoral research fellow with Data Communications
Group, Department of Electrical and Computer
Engineering, University of British Columbia. He is now a postdoctoral research
fellow at King Abdullah University of Science and Technology (KAUST),
Thuwal, Mekkah Province, Saudi Arabia. His research interests include:
performance limits of communication systems, cognitive and sensor networks,
physical-layer security, and low-complexity detection algorithms.
\end{biography}

\begin{biography}{Mohamed-Slim Alouini} (S'94, M'98, SM'03, F'09) was
born in Tunis, Tunisia. He received the Ph.D. degree in electrical engineering
from the California Institute of Technology (Caltech), Pasadena,
CA, USA, in 1998. He was with the department of
Electrical and Computer Engineering of the University of Minnesota,
Minneapolis, MN, USA, then with the Electrical and Computer Engineering
Program at the Texas A$\&$M University at Qatar,
Education City, Doha, Qatar. Since June 2009, he has been a Professor
of Electrical Engineering in the Division of
Physical Sciences and Engineering at KAUST, Saudi
Arabia., where his current research interests include the design and
performance analysis of wireless communication systems.
\end{biography}

\end{document}